\documentstyle[secnumtab]{fbssuppl}
\title{Recent Developments in one and two Pion Production in Elementary
Reactions and Few Body Systems
\\
(Talk given at 'XVth European Conference on Few-Body Problems in Physics,
FB XV)
}
\author{E. Oset, F. Cano, J.A. G\'omez Tejedor, E. Hern\'andez and
M. J. Vicente Vacas}
\institute{ Departamento de F\'{\i}sica Te\'orica and IFIC,
Centro Mixto Universidad de Valencia-CSIC,
46100 Burjassot (Valencia), Spain}

\input{psfig.sty}

\begin{document}
\maketitle

\begin{abstract}
In this talk we cover several issues concerning pion production. The first
one is the $p p \rightarrow p p \pi^0$ reaction where an alternative
explanation
based on the off shell extrapolation of the $\pi N$ amplitude is offered.
A recent model for the $\gamma N \rightarrow \pi \pi N $ reaction is presented
and a new kind of exchange current is constructed from it which allows one
to address problems like double $\rmDelta$ photoproduction from the deuteron.
Finally the $(\gamma, \pi \pi)$ reaction in nuclei is studied and shown to
be highly sensitive to the renormalization of the pions in nuclei.
\end{abstract}

\section{The $p p \rightarrow p p \pi^0$ reaction at threshold.}

This reaction has been addressed in different works \cite{1,2} and
shown to be in strong disagreement with very precise data obtained at the
Indiana Cyclotron \cite{3}. The starting point is a model which contains the
impulse approximation, diagram 1a) and the rescattering term, diagram 1b).

\begin{figure}
\centerline{\protect\hbox{\psfig{file=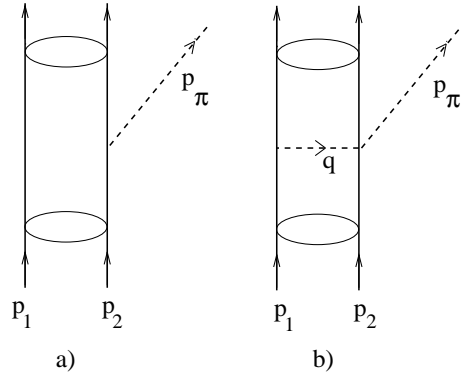,width=.5\textwidth}}}
\caption{Feynman diagrams considered in the $p p \protect\rightarrow p p
\protect\pi^0$ reaction near threshold. a) Born term; b) rescattering term.
}
\end{figure}

The first mechanism involves the usual Yukawa vertex and is only possible
due to the initial and final state interaction. The second one involves the
$\pi N$ scattering amplitude in one of the vertices which is given in terms
of an effective Lagrangian

\begin{equation}
\delta H = 4 \pi \frac{\lambda_1}{m_{\pi}}
\bar{\psi} \vec{\phi} \vec{\phi} \psi + 4 \pi \frac{\lambda_2}{m_{\pi}^2}
\bar{\psi} \vec{\tau} \vec{\phi} \times \dot{\vec{\phi}} \psi
\end{equation}

\noindent
where $\lambda_1 = 0.0075$ and $\lambda_2 = 0.053$
for on shell pions at threshold. The $\phi_3 \times \dot{\phi}_3$
combination is null and hence the second term in eq. (1) does not contribute
for the $\pi^0 p \rightarrow \pi^0 p$ amplitude.

Hence only the $\lambda_1$ term contributes, which gives a small contribution
given the smallness of $\lambda_1$. Actually $\lambda_1$ is strictly zero
in the limit of exact chiral symmetry. A way out of the puzzle was offered in
\cite{4} in terms of a relativistic exchange current involving $\sigma$
exchange
between positive and negative energy components. This exchange is tied to the
scalar potential, $V_{s}$, of the relativistic potential of Walecka type
$V_{S} + \gamma^{0} V_{V}$.
This interpretation which can not be excluded a priori,
does not have however a strong base, because the exchange current is generated
from a relativistic form of the $N N \rightarrow N N$ amplitude which is
unique for on shell nucleons. However, there are an infinite amount of possible
forms of the amplitude which are equivalent for on shell nucleons but which
provide quite different off shell extrapolations
or extensions to the states of
negative energy, as is the case in \cite{4}. One typical
example is the $\gamma _5$ matrix which is equivalent
to  $ q \! \! \! \slash \, \gamma_5 / 2 m$ for on shell nucleons but which
provides different extensions in the negative energy sector.
The work of \cite{4} is based on a particular
choice for the $N N$ amplitude but one can obtain
quite different results for the $p p \rightarrow p p \pi^0$ reaction if
one starts from other expressions which are equivalent
for on shell nucleons. More concretely, recent models \cite{5}
which consider $\sigma$ exchange
as a correlated two pion exchange would
provide a different answer in the negative energy sector
than the ordinary $\sigma$ exchange.

We have followed here a different idea
suggested in \cite{6}. The idea is that the $\pi N$
amplitude appearing in the rescattering term,
fig. 1b) is not the on shell $\pi N$ amplitude but
the half off shell amplitude and this one is much
bigger than the on shell one in all
existing models of the off shell extrapolation.

\begin{figure}
\centerline{\protect\hbox{\psfig{file=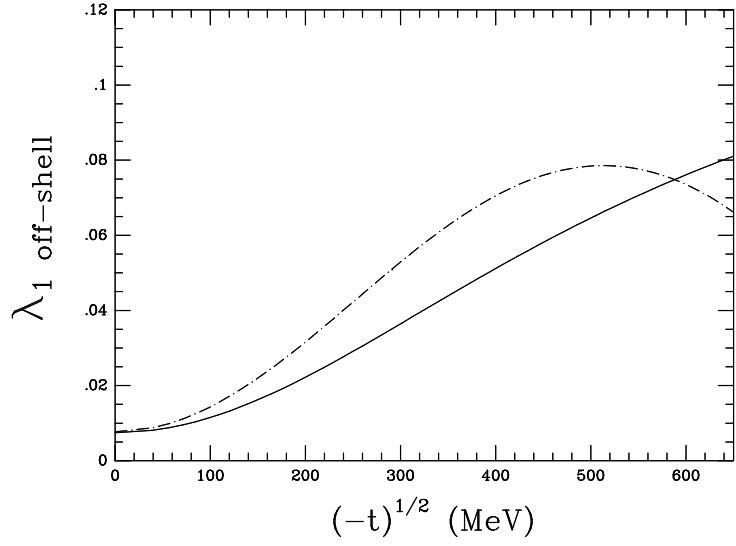,width=.8\textwidth}}}
\caption{Off shell extrapolation of the $\protect\pi N$ isoscalar amplitudes
in the Hamilton \protect\cite{8} (solid line) and current algebra model
\protect\cite{9} (dashed-dotted line).
}
\end{figure}

\begin{figure}
\centerline{\protect\hbox{\psfig{file=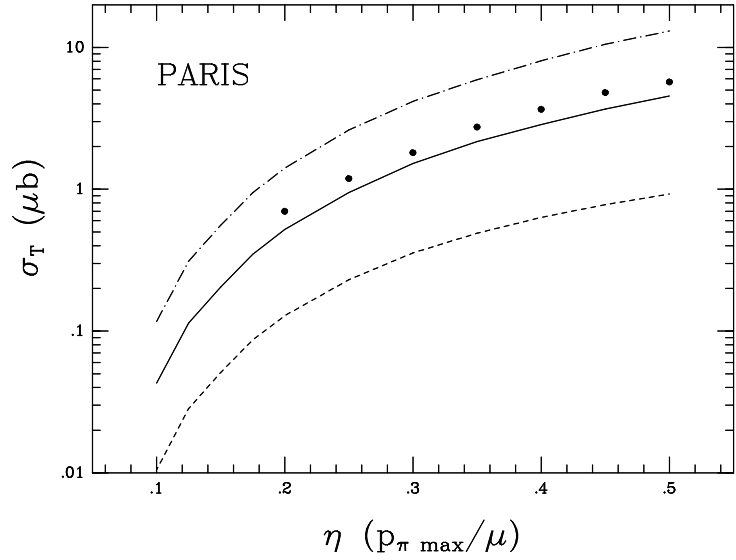,width=.8\textwidth}}}
\caption{Cross section for $p p \rightarrow p p \protect\pi^0$ near
threshold using the Paris $N N$ potential for the initial and final state
interaction. Dashed line: with in shell $\protect\pi N$ isoscalar amplitude.
Solid line: with off shell $\protect\pi N$ amplitude from Hamilton's model
\protect\cite{8}. Dashed dotted line: same with the current algebra
$\protect\pi N$ extrapolation \protect\cite{9}.
}
\end{figure}

In a recent work \cite{7} we have carried out
detailed calculations of the $p p \rightarrow p p \pi^0$ cross section
using two different off shell extrapolations for $\lambda_1$, one
from Hamilton \cite{8} which contains
a $\sigma$ exchange with the $\sigma \pi \pi$ and $\sigma N N$ vertices
and a short range piece, which approximately cancel
on shell. The other one, from Banerjee \cite{9},
is based on PCAC and has also been extensively used
(we use a regularized version of it to have a well behaved Fourier Transform).
The two curves can be seen in fig. 2, where we
appreciate the large enhancement of the off shell
amplitude for momenta around 400 MeV which
are the relevant ones for the present problem
(the upper curve corresponds to \cite{9} and the lower one to
\cite{8}).

In fig. 3 we show the results which we
obtain for the cross section of the $p p \rightarrow p p \pi^0$
reaction close to threshold. The dashed
curve is the result obtained using the $\pi N \rightarrow \pi N$
on shell amplitude, while the dash-dotted curve and the solid line give the
results obtained using the Banerjee and
Hamilton extrapolations respectively. The
results are obtained using the Paris potential
for the initial and final state interaction and
the results are similar, although a little
higher, if the Bonn potential is used.

It is clear from the results that one can not be very assertive, because
there are still uncertainties on the off shell extrapolation,
but the results plotted there clearly show
that the off shell extrapolation by itself
can easily explain the experimental data,
offering a down to earth explanation of the
discrepancies with experiment based on the use of the on shell $\pi N$
amplitude.

\section{Double pion photoproduction on the nucleon.}

A recent model for the $\gamma p \rightarrow \pi^+ \pi^- p$ reaction has
been presented in \cite{10}. The model is
schematically depicted in fig. 4, where the baryonic intermediate states
stand for $N, \rmDelta, N^* (1440)$ and $N^* (1520)$. In total 67 Feynman
diagrams appear when this states are considered, although some terms found
to be negligible are not taken into account. At energies below $E_{\gamma}
= 800 $ MeV many of those terms are also negligible and a simplified model,
fig. 5, which contains only 20 terms, is sufficiently good, and this is the one
used for the study of the other isospin channels.

\begin{figure}
\centerline{\protect\hbox{\psfig{file=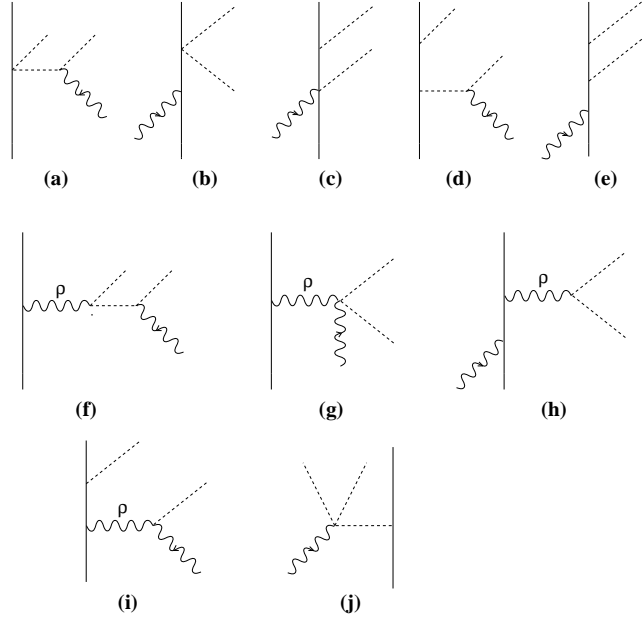,width=.7\textwidth}}}
\caption{Classification of the Feynman diagrams into one point, two point and
       three point diagrams.
       Continuous straight lines: baryons.
       Dashed lines: pions.
       Wavy lines: photons and $\protect\rho$-mesons (marked explicitly).
}
\end{figure}

\begin{figure}
\centerline{\protect\hbox{\psfig{file=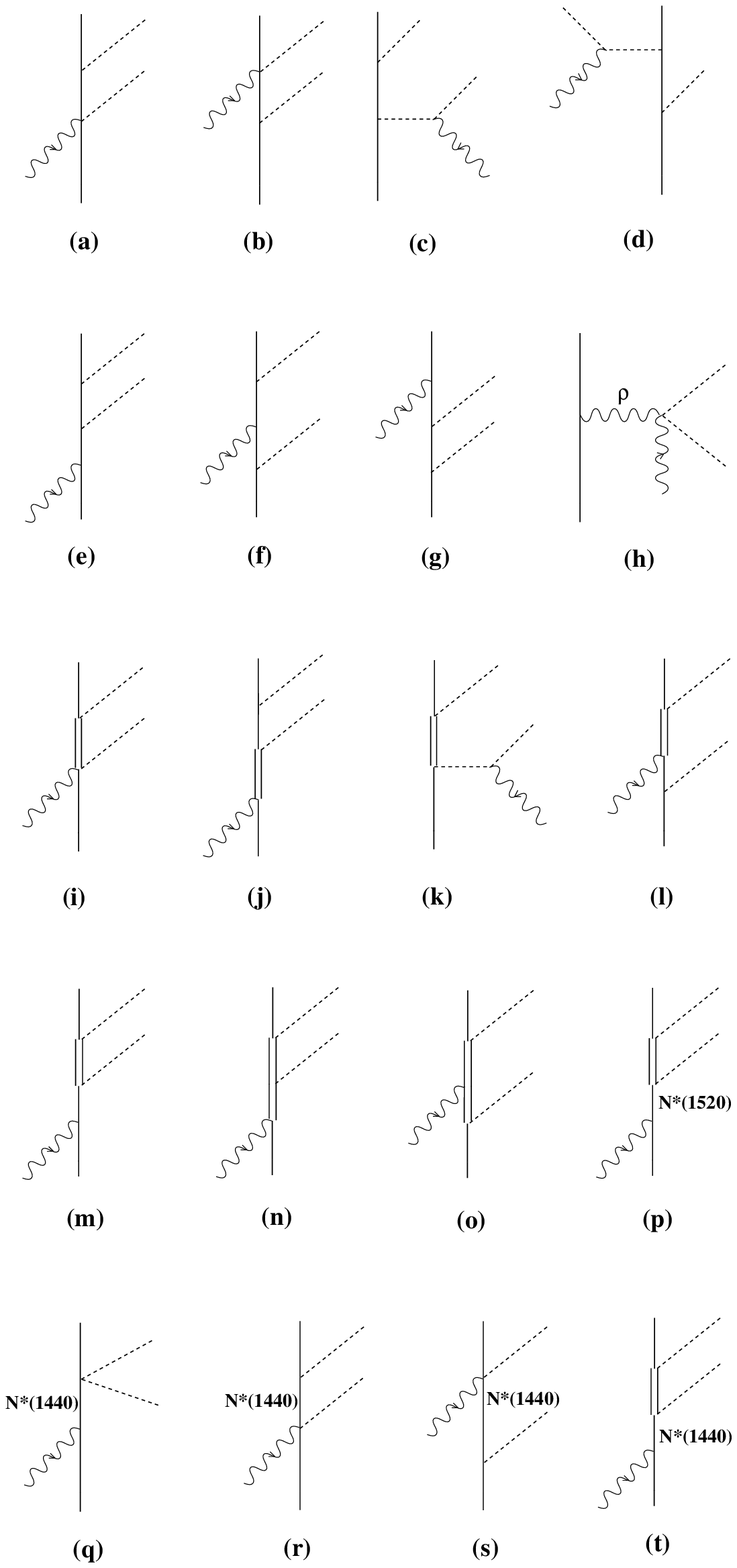,width=.8\textwidth}}}
\caption{ Feynman diagrams for the
      $\protect\gamma N \protect\rightarrow \protect\pi \protect\pi N$
      reactions.
}
\end{figure}

The results for the $\gamma p \rightarrow \pi^+ \pi^- p$ reaction are
shown in fig. 6. Four different lines are depicted there which we
explain below.

\begin{figure}
\centerline{\protect\hbox{\psfig{file=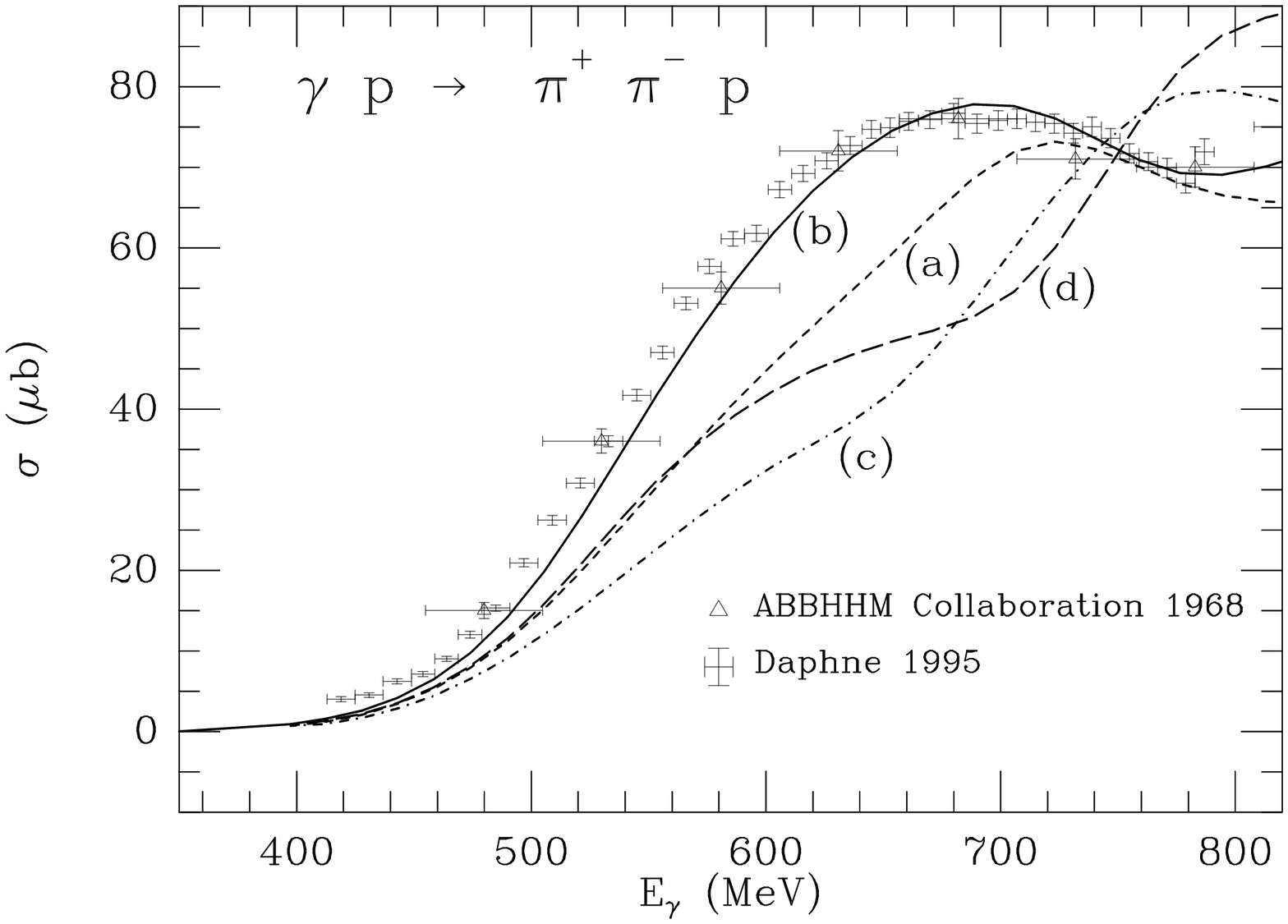,width=.9\textwidth}}}
\caption{ Total cross section for the $\protect\gamma p \protect\rightarrow
       \protect\pi^+ \protect\pi^- p$ reaction. See text for explanation.
}
\end{figure}

The dominant term in the $\gamma p \rightarrow \pi^+ \pi^- p$
amplitude is the one involving the $\gamma N \rmDelta \pi$ Kroll-Ruderman
term (diagram (i) in fig. 5).
It just happens that the amplitude depicted in fig. 5(p), with the excitation
of the $N^* (1520)$ resonance which later on decays
in $\rmDelta^{+ +} \pi^{-}$,
has a piece in the amplitude with exactly the same spin and momentum structure
as the dominant
term of fig. 5(i) and gives rise to a strong
interference even if individually it is a small term. This is one of the
surprising and interesting accidents of this reaction which allows one to
determine the signs of the $s$- and $d$-wave amplitudes for the
$N^* (1520) \rightarrow \rmDelta \pi$ reaction, something which
poses new challenges to quark models of the hadrons \cite {11}. To envisage
this we write here the basic ingredients of these amplitudes which are the
couplings $N^* (1520) \rightarrow N \gamma$ and
$N^* (1520) \rightarrow \rmDelta \pi$:

$$
-i \delta H_{N^* N \gamma} =
i g_{\gamma} \vec{S}^{\dagger} \cdot \vec{\varepsilon}
+ g_{\sigma} (\vec{\sigma} \times \vec{S}) \cdot \vec{\varepsilon}
$$

\begin{equation}
-i \delta H_{N^* \rmDelta \pi} = - [ f + \frac{g}{\mu^2} (\vec{S}_{\rmDelta}
\cdot \vec{q})^2] T^{\dagger \lambda}
\end{equation}

\noindent
where $\vec{S}, \vec{T}$ are spin, isospin transition operators from spin,
isospin 1/2 to 3/2, $\vec{S}_{\rmDelta}$ the ordinary spin matrix for spin 3/2
and $\vec{\epsilon}$ is the photon polarization vector. The amplitudes
$g_{\gamma}$ and $g_{\sigma}$ are fitted to the experimental $N^* (1520)$
helicity amplitudes,
and the strength of the $s$-wave and
$d$-wave partial widths in the $N^* \rightarrow
\rmDelta \pi$ decay are also taken from experiment. This gives, however,
four possible solutions:

$$a)\; f = -0.19 \; , \; g = 0.18 \quad ; \quad b) \; f = 1.03
\; , \; g = - 0.18 \; ; $$
\begin{equation}
c)\; f = -1.03 \; , \; g = 0.18 \quad ; \quad d) \; f = 0.19
\; , \; g = - 0.18 \;
\end{equation}

The interference of the amplitudes is between the Kroll-Ruderman term of fig.
5(i), and the $N^*$ term of fig. 5(p) through the $s$-wave decay amplitude of
the  $N^* \rightarrow \rmDelta \pi$. Thus one finds a combined amplitude

\begin{equation}
i \frac{f^*}{\mu} \vec{S} \, \cdot \, \vec{q}_+ D_{\rmDelta} \left[ e
\frac{f^*}{\mu} -
\left(g_{\gamma} - g_{\sigma}\right) \left( f + \frac{5}{4} g \frac{\vec{q}\,
^{2}}{\mu^2} \right) D_{N^*} \right] \vec{S}^{\dagger} \cdot \vec{\varepsilon}
\end{equation}

\noindent
where $\vec{q}_{+}$ is the momentum of the $\pi^{+}$ and $D_{\rmDelta},
D_{N^*}$ are the propagators of the $\rmDelta$ and $N^*(1520)$ respectively.

One can see in fig. 6 that out of the four possible solutions of eq. (3)
only one, solution (b), gives rise to good agreement with experiment \cite{12}
{}.
The reason of the interference can be seen in eq. (4). Since
$g_{\gamma} - g_{\sigma} = 0.157 > 0$ and
$( f + \frac{5}{4} g \vec{q}\, ^{2} / \mu^2) > 0$
one finds constructive interference below the $N^* (1520)$ energy and
destructive
interference above that energy and this interference is what gives rise to
the peak in the cross section, which does not appear without the $N^* (1520)$
contribution.

\begin{figure}
\centerline{\protect\hbox{\psfig{file=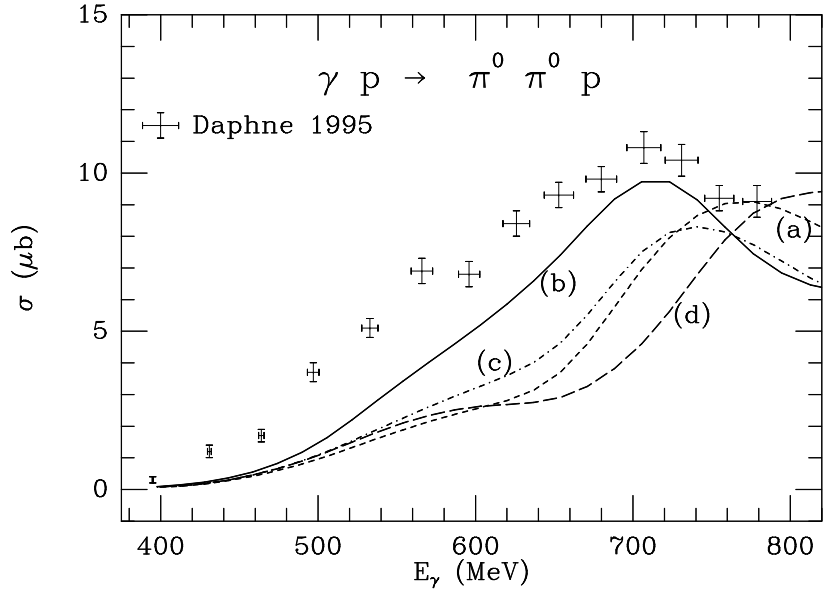,width=.9\textwidth}}}
\caption{Total cross section for the $\protect\gamma p \protect\rightarrow
 \protect\pi^0 \protect\pi^0 p$ reaction. See text for explanation.
}
\end{figure}

We have calculated the cross section for other channels. One of them,
particularly interesting, is the $\gamma p \rightarrow \pi^0 \pi^0 p$
also measured in \cite{12}. We can see in fig. 7 that again only the solution
(b) is the one that best fits the data. We should mention that the TAPS
collaboration has also preliminary data \cite{13} which agree even better
with our results in this channel.

For reasons of space we shall not show here the results for other channels.
Only we mention that we find disagreement with the experimental results
in the $\gamma p \rightarrow \pi^+ \pi^0 n$ channel for which we do not
yet envisage an explanation.

\section{Double $\rmDelta$ production in the $\gamma d \rightarrow p n
\pi^+ \pi^-$ reaction}

In a recent paper \cite{14} we have tackled the reaction $\gamma d \rightarrow
\rmDelta^{+ +} \rmDelta^{-}$ which has been measured recently \cite{15}.

The excitation of two $\rmDelta$ necessarily requires the participation of two
nucleons. We need some exchange currents mechanism which generate this
transition. The previous model for the $\gamma p \rightarrow \pi^+ \pi^- p$
reaction offers us a clue on how to proceed. Let us take the
 diagrams of fig. 5(i), 5(p), which together with the pion pole term
 (diagram (k) of fig. 5),
give already a fair description of the reaction. We can attach the $\pi^-$
produced to a second nucleon and produce a second $\rmDelta$. The diagrams
which appear are shown in fig. 8, providing a fair model for this reaction
up to energies of around $E_{\gamma} \simeq 800 $ MeV, where the elementary
model for the $\gamma p \rightarrow \pi^+ \pi^- p$ reaction starts having
deficiencies.

\begin{figure}
\centerline{\protect\hbox{\psfig{file=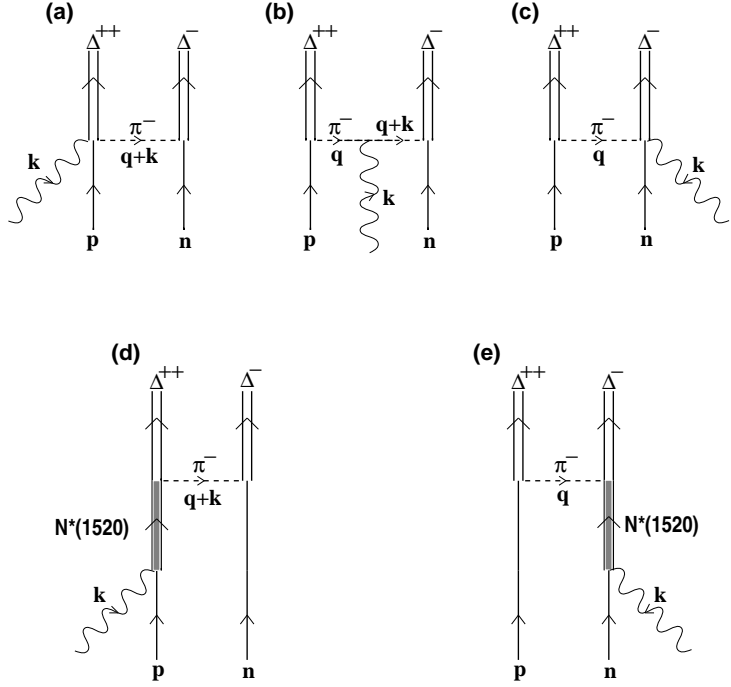,width=.8\textwidth}}}
\caption{Terms considered in our model for the $\gamma d \rightarrow
\Delta^{++} \Delta^{-}$ reaction.
}
\end{figure}

The results for the $\gamma d \rightarrow \rmDelta^{+ +} \rmDelta^{-}$ reaction
can be seen in fig. 9 compared with the experiment. The dashed line omits
the contribution of the $N^* (1520)$ while the solid one takes it into account.
We can see that there is a good agreement with experiment up to around
$E_{\gamma} = 800$ MeV and the data hint at discrepancies from there on.
The consideration of the $N^* (1520)$ term seems to improve the results a bit.

\begin{figure}
\centerline{\protect\hbox{\psfig{file=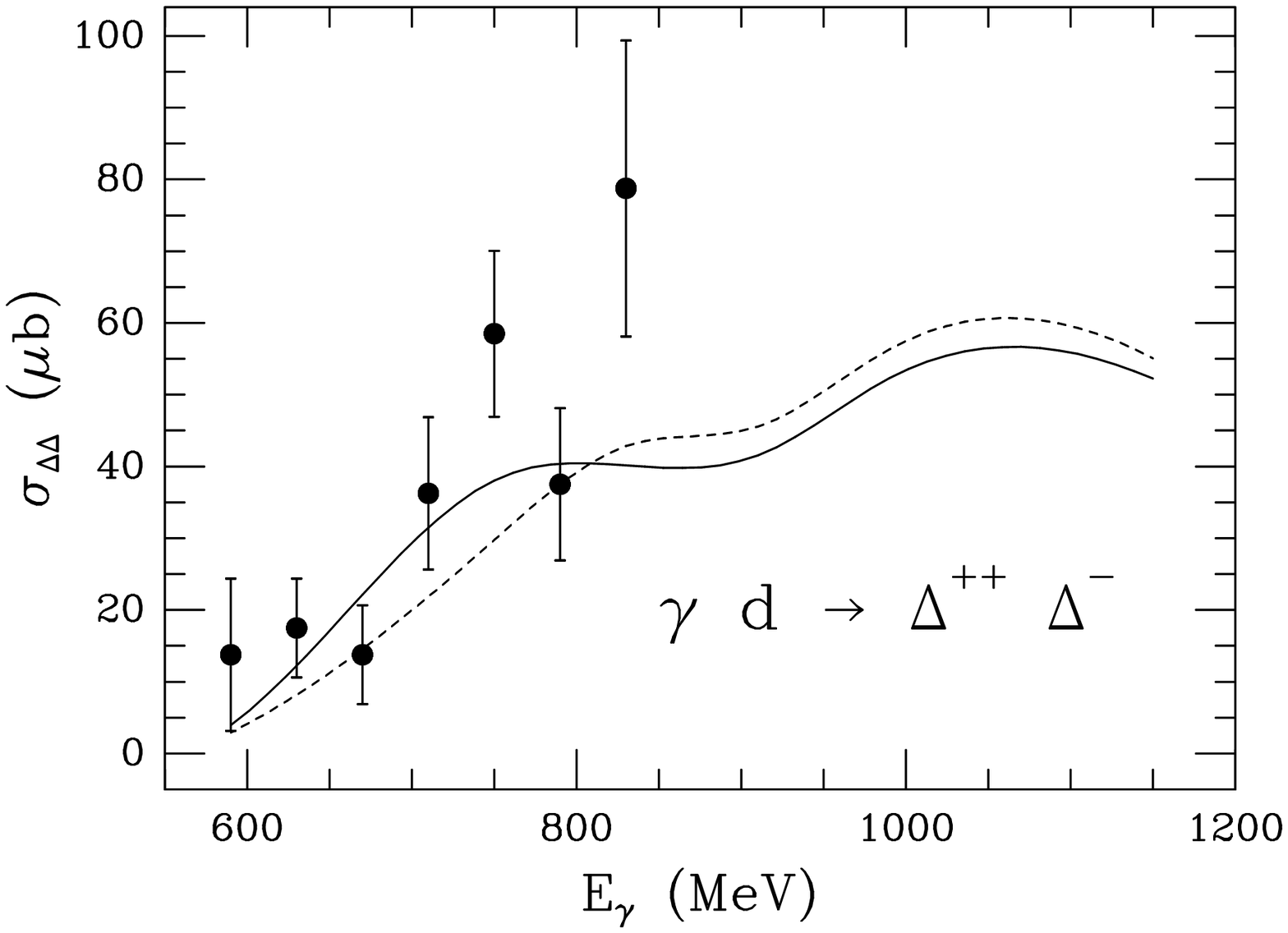,width=1\textwidth}}}
\caption{Results of the model compared to the data of ref.
{\protect\cite{15}}.
Dashed line, omitting the $N^{*}$ terms. Solid line,
results including
all terms of the model of Fig. 8.
}
\end{figure}

\section{Double pion photoproduction in nuclei.}

We have studied the $(\gamma, \pi^+ \pi^-)$ inclusive reaction in nuclei
in order to see renormalization effects due to the interaction of pions
with the nucleus. Such effects were predicted in \cite{16} and observed in
\cite{17} for the $(\pi, 2 \pi)$ reaction in nuclei and they could only be
magnified here since the photons penetrate deeper than the pions. We evaluate
the cross section by assuming the nucleus to be made of bits of infinite
nuclear matter at each point of the nucleus. This is the essence of the local
density approximation, which given the fact that the photons explore the
whole nuclear volume, is a very good approximation \cite{18}.

The cross section for this process in the ``improved''impulse approximation
would be

$$ \sigma = \frac{\pi}{k} \int d^3 r \int \frac{d^3 p}{(2 \pi)^3}
\int \frac{d^3 q_1}{(2 \pi)^3} \int \frac{d^3 q_2}{(2 \pi)^3}
\sum_{\alpha} \overline{\sum} \sum |T_{\alpha}|^2 $$

$$n_{\alpha} (\vec{p}) [ 1 - n_{\alpha} (\vec{k} + \vec{p} - \vec{q}_1 -
\vec{q}_2)] \frac{1}{2 \omega (q_1)} \frac{1}{2 \omega (q_2)} $$

\begin{equation}
2 \pi \delta (k^0 + E (p) - \omega (\vec{q_1}) - \omega (\vec{q_2}) - E
(\vec{k} + \vec{p} - \vec{q}_1 - \vec{q}_2))
\end{equation}

\noindent
where $\vec{p}$ is the momentum of the nucleons from the Fermi sea, $\vec{k}$
the momentum of the photon and $\vec{q}_1, \vec{q}_2, \omega (\vec{q}_1),
\omega (\vec{q}_2)$ the momenta and energies of the two pions. The variable
$\alpha$ indicates isospin channels, and $n_{\alpha} (\vec{p})$ is the
nucleon occupation number in the Fermi sea. Since

\begin{equation}
2 \int \frac{d^3 p}{(2 \pi)^3} \; n_{\alpha} (\vec{p}) = N \; \hbox{or} \; Z
\end{equation}

\noindent
one is essentially multiplying by $N$ or $Z$ (and summing) the individual
$\gamma N \rightarrow \pi^+ \pi^- N $ cross sections, with two important
modifications: i) the Fermi motion is explicitly taken into account
ii) the Pauli blocking factor $[ 1 - n ]$ is explicitly considered there.
The formulation of the problem is easily done in terms of the selfenergy of
the photon, $\Pi (\vec{k})$ and the relationship

\begin{equation}
\sigma = - \frac{1}{k} \int Im \Pi \left(\vec{k}, \rho \left(\vec{r}\right)
\right) d^3 r
\end{equation}

\noindent
which leads to eq. (5). The formulation in terms of a selfenergy diagram
allows for systematic corrections. An important one arises from the
consideration of the renormalization of the pion propagators, which essentially
substitutes $\omega ({\vec q})$ in eq. (5) by $\tilde{\omega} (q)$, the pion
dispersion
relation in the medium. Simultaneously we also renormalize the $\rmDelta$ and
pion propagators in the amplitudes for the model of $(\gamma, \pi^+ \pi^-)$.
In addition one must distort the pion waves by the imaginary part of the
pion selfenergy due to absorption in order to take into account the absorption
of the pions in their way out of the nucleus.

\begin{figure}
\centerline{\protect\hbox{\psfig{file=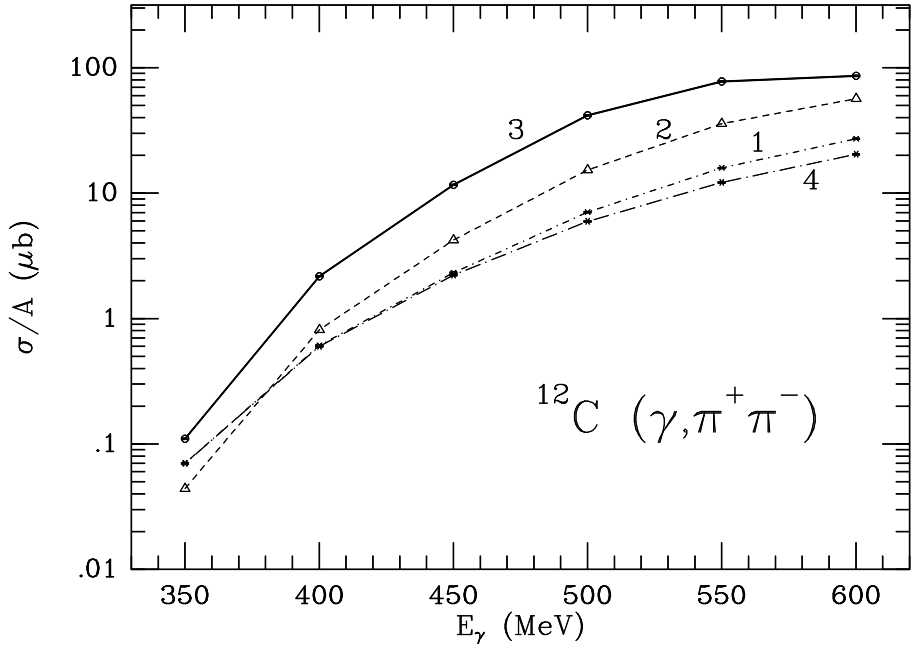,width=1\textwidth}}}
\caption{Cross section per nucleon for
the $ ^{12}C (\gamma, \pi^+ \pi^-)$ reaction.
Dot-dashed line (labelled 1): Simple impulse approximation.
Dashed line (labelled 2): Scaling of the deuteron cross section.
Continuous line (labelled 3): Full model.
Long dashed-dotted curve (labelled 4): cross section renormalizing
the internal pion and delta lines in the amplitudes
and ignoring the renormalization
of the external pion lines.
}
\end{figure}

In fig. 10 we can see the results of the evaluation of \cite{19}.
The curve labelled 1 is the impulse approximation
$N \sigma_n + Z \sigma_p$ multiplied by the pion absorption factor.
The curve 2 is a pure scaling of the cross section in the deuteron
$(A \sigma_d / 2)$, while curve 3 is the result from our calculations. The
differences between 1 and 3 indicate effects of renormalization, which increase
the cross section by a factor from 4 to 5. It is also interesting to
see that in spite of the absorption of the final pions, the cross section
obtained is still larger than A/2 times the cross section in the deuteron.

It would be interesting to perform such experiments, easily implementable at
Mainz as a complement to the $(\gamma, \pi \pi)$ program on the nucleon, and
in general to the whole program of reaction mechanisms in photonuclear
interactions.

\section{Summary:}

We made a short review of reactions producing one or two pions in elementary
processes, in deuteron and $^{12}C$. The off shell dependence of the isoscalar
$\pi N$ amplitude was shown to be very important in the $p p \rightarrow
p p \pi^0$ reaction. On the other hand a rather complete model for the
$\gamma N \rightarrow \pi \pi N$ reaction is made, which is bound to play a
similar role as the $\gamma N \rightarrow \pi N$ reaction when the energy
of the experiments go up to the GeV regime as in Mainz. We offered an example
of exchange currents produced from that model in the $\gamma d \rightarrow
\rmDelta^{+ +} \rmDelta^-$ reaction. Finally we also showed important
renormalization
effects in the $(\gamma, \pi^+ \pi^-)$ reaction in nuclei, which once more
stresses the role played by pion physics in photonuclear reactions in general.

\end{document}